\newcommand{\be}{\begin{equation}}
\newcommand{\en}{\end{equation}}
\newcommand{\del}{\delta}

\documentstyle[12pt]{article}
\begin{document}
\title{ Correlated Wave Packet treatment of Neutrino
and Neutral Meson  Oscillations}
\date{Nov. 1998}
\author{Michael Nauenberg\\
Department of Physics\\
University of California, Santa Cruz, 
}
\maketitle

\begin{abstract}

A quantum wave packet treatment of neutrino 
and neutral K and B meson oscillations
is presented  which incorporates the recoil  
particle in the production process, and includes the effect of   
the localization  and lifetime of the source
assumed to be a resonance or unstable particle.
This approach removes  the ambiguities in the conventional 
single particle 
treatment of these oscillations, and elucidates the role of
quantum correlations with the recoil particle. 
A fundamental connection between the stochastic decay time of the source
and the space-time coordinates of the correlated final state particles 
is derived.

\end{abstract}

A proper description of neutral K  and 
B meson oscillations  \cite{bell1} \cite {bmeson1}  
and  neutrino oscillations \cite {lsnd1}-\cite {kamiokande1} 
requires that the familiar superposition of states with  definite mass 
be represented by coherent wave packets
\cite {nussinov1} - \cite {pevsner}.
However, the conventional {\it single} particle treatment of these oscillations 
leads to  ambiguities which have lead to debates  whether 
the momentum \cite {nussinov1} - \cite {kayser1} or the 
energy  \cite {lipkin1} -\cite {lipkin4}  remains unchanged
for different  mass eigenstates, 
although  the resultant transition probabilities 
are the same in both cases. 
Furthermore, such descriptions leave  unanswered the 
fundamental question how the properties of a single particle 
wave packet are determined by the nature of the  production
process. In this note we resolve these problems by considering a 
a wave packet which includes a single  recoil
particle produced by the decay of a resonance or unstable
particle, and incorporates both energy and momentum conservation \cite {giunti1}  
-\cite {okun1} instead of 
resorting to conventional  ad-hoc assumptions.
Including final state correlations raises  some  interesting new questions
because these correlations take place 
between  space-like separated events,  a subtle problem in quantum
mechanics  discussed  a long time ago by Einstein, Podolsky and Rosen (EPR) 
 \cite {epr1}, and  considered in connection with the production 
of a neutral $K \bar  K$  or $B \bar B$ pair in \cite {lipkin3} - \cite {rosner}.
In this case oscillations can be observed  as EPR correlations
between  neutral mesons of fixed flavor  or their decay products. 
It has been claimed in  \cite {widom0} - \cite{widom3} 
that  oscillations  of the  recoil particle 
can also  be observed  even if this particle  has a fixed mass. 
In the case of neutrinos produced in  the decay $\pi \rightarrow \mu+ \nu$
where  the recoil is  a  charged lepton, this would  greatly facilitate
experiments.
However,  Lowe et al. \cite {goldman1}-\cite {goldman2} have argued  
that  only a certain traveling pattern of oscillations 
in the recoil particle coordinates  is observable,
while Dolgov et al. \cite {okun1} concluded  that such  
oscillations could be observed  only as EPR correlations  provided that
the detection of the recoil lepton is related to a  neutrino
of  fixed  flavor.

These problems are examined here in a quantum mechanical wave packet treatment 
of neutrino and neutral meson  oscillations which includes  the
recoil particle in an entangled state.  We make explicit
certain assumptions concerning  correlations
between two particles which  have  space-like separations.
An important  feature in this  approach is that a  correlated wave packet
can incorporate  the effect of both  localization and finite lifetime of
the source, which is assumed here to be a resonance or unstable particle.
It will be shown that  these  properties explain
why the propagation of the particles  is confined  near classical trajectories.
While it has been recognized  that classical motion must be combined with
wave properties and interference effects  for an understanding of the oscillation phenomena,
in current discussions classical trajectories  
have not been introduced in a self-consistent manner
\cite {okun1}, \cite {widom0} - \cite {goldman2}. 
A  novel  property of our wave packet
is that it can  incorporate  the space-time coordinates of the 
decay point of the initial unstable state
which can be observed and provides  a reference point for the oscillations. 
Our results averaged over unobserved recoil particle coordinates, Eqs. \ref {avcost}
and \ref {avcosx}, 
differ from those of Dolgov et. al. \cite {okun1}, who
assumed that the source and  decay particles satisfy exact classical equations motion
in violaton of the principles of quantum mechanics. 

For simplicity we consider the theory in one dimension where all the processes
are collinear, and assume there are only two mass eigenstates $|a>$ and $|b>$  
which is adequate  for our purposes. 
Then the  transition amplitude to some final state $|g>$ is
\be
\label {amplitude1}
A \propto  cos(\theta) <g|a>\psi_a + sin(\theta)<g|b> \psi_b 
\en
where $\theta$ is the mixing angle for an initial state of definite 
flavor or strangeness, and $\psi_a$ and $\psi_b$ are wavefunctions 
associated with the different  mass eigenstates. 
We  obtain these wavefunctions  by time-dependent
perturbation theory, assuming that  the initial state is a resonance or unstable state 
of mass $M$ and width $\Gamma$. The  wavefunction  for this state for $t \geq 0$  is
\be
\label{init1}
\psi_o (x,t)=\int dp f(p) exp[i p (x-x_s) -(iE_p+M\Gamma/2E_p)(t-t_s)]
\en
where $t_s$ is the time at which this state is created as a wave packet centered at $x_s$,  
$f(p)$ is the  amplitude associated with a momentum distribution  $p$  
in the initial state with corresponding energy $E_p=\sqrt{p^2+M^2}$. 
In the following discussion  we set for convenience $x_s,t_s$ at the 
origin of our space-time coordinate system , but
it should be remembered that in practice these coordinates are not known precisely
and will be included in some of our main results.
We assume that this amplitude has a sharp maximum at $p=\bar p$,  
and expanding $ E_p$ to first order in $p-\bar p$ we obtain
\be
\label{gaussian1}
\psi_o(x,t) = exp [i(\bar p x - E_{\bar p}t )] ~ exp(-M\Gamma t/2E_{\bar p})~g(x-\bar v t)
\en
where the envelope of the wave packet is given directly by the wavefunction
of the source at $t=0$
\be
\label{envelope1}
g(x)=exp(-i\bar p x)\psi(x,0)
\en
In the Wigner-Weisskopf approximation we obtain   for $ t \geq 0$ 
\be
\label {psi1}
\psi_{a,b}(x_1,x_2,t)= N  \int dp_1\int dp_2  f(p)  \frac{exp(ip_1 x_1+ip_2 x_2-i(E_1+E_2)t)}{(E_1+E_2-E_p+i M\Gamma/2E_p)}
\en
where $N=(1/2\pi)\sqrt{\Gamma M |\bar v_{12}|/E_{\bar p}}$, $\bar v_{12}$ is the mean relative
velocity, and
$E_1=\sqrt{p_1^2+m_1^2}$ and  $E_2=\sqrt{p_2^2+m_2^2}$ are the 
relativistic energies of the two correlated particles in the final state 
with masses  $m_1$ and $m_2$ which can  have different values  for the eigenstates
labeled $a$ an $b$.   
Conservation of total momentum in the 
production process implies that 
\be
\label {mom1}
p = p_1+p_2,
\en

A shortcoming of this representation for $\psi$  is that  
the state of both particles is  given  at the same time $t$,
while in practice these particles can be  detected at different times $t_1$
and $t_2$. However, since these particles are not  interacting,  the subsequent time evolution 
of the wavefunction can be determined by their respective free particle Hamiltonians 
$H_1$ and $H_2$. Hence
\be
\label {psi11}
\psi(x_1,x_2,t_1,t_2)=exp(-iH_1(t_1-t)-iH_2 (t_2-t))\psi(x_1,x_2,t),
\en
and  the required wavefunction  $\psi (x_1,x_2,t_1,t_2)$ (footnote 1) is
obtained by replacing the factor $(E_1+E_2)t$ in Eq. \ref {psi1} by
$E_1 t_1+E_2 t_2$. If these particles are unstable, as is the case 
for neutral kaons,  an additional factor 
in the integrand of Eq. \ref{psi1} is required, of the form  
\be
\label{lifetimes1}
exp-(\frac{m_1 t_1}{2 \tau_1  E_1}+\frac{m_2 t_2}{2 \tau_ 2 E_2}),
\en 
where $\tau_1$ and $\tau_2 $ are the particle lifetimes. 
The assumption that the  initial state is a resonance 
or unstable particle of width $\Gamma$  
implies an uncertainty in the total
energy which plays an essential  role in understanding how  
the decay  particles are confined to classical trajectories.
Nevertheless,  we can define 
mean momenta  $\bar p_1$ and $\bar p_2$ associated with
the  mean total momentum $\bar p$  by the requirement that for these
special values of momenta, the energy conservation
relation
\be
\label {en1}
E_{\bar p}= \bar E_1+ \bar E_2.
\en
is satisfied exactly.

We carry out the integrations in Eq. \ref{psi1} approximately, 
by expanding the momenta $p_1$ and $p_2$  around these mean values  $\bar p_1$ and $\bar p_2$,
obtained as solutions of the momentum- energy conservation equations,  
Eqs. \ref{mom1} and  \ref{en1} to first order in $p-\bar p$ and  $E- E_{\bar p}$, 
where $E=E_1+E_2$. Second order terms contribute to the dispersion of the
wave packet which we neglect here.  We have
\be
p_1=\bar p_1 +\frac{ (E- E_{\bar p})-\bar v_2 (p-\bar p)}{ \bar v_{12}}, 
\en
and
\be
p_2=\bar p_2 -\frac{ (E- E_{\bar p})-\bar v_1 (p-\bar p)}{ \bar v_{12}}, 
\en
where $\bar v_1=\bar p_1/\bar E_1$, $\bar v_2=\bar p_2/\bar E_2$ are 
the classical or group velocities, and  $ \bar v_{12} = \bar v_1-\bar v_2$ is the relative
velocity of particles $1$ and $2$. Likewise
\be
E_1 = \bar E_1 + \bar v_1 (p_1-\bar p_1),
\en
and
\be
E_2 = \bar E_2 + \bar v_2 (p_2-\bar p_2).
\en
Changing the coordinates in the integrand of Eq. \ref {psi1} to 
the variables $p$ and $E$, we obtain 
for $t_{12}>0$
\be
\label{psi12}
\psi(x_1,x_2,t_1,t_2) = N'exp[i\phi_{12}]~exp(-M\Gamma t_{12}/2E_{\bar p})g(z_{12}), 
\en
where $N'=-i\sqrt{\Gamma M/E_{\bar p} \bar v_{12}}$ is a constant, and
\be
\phi_{12} = \bar p_1 x_1 + \bar p_2 x_2-\bar E_1 t_1-\bar E_2 t_2, 
\en
\be
t_{12}=(\Delta x_2 -\Delta x_1)/\bar v_{12},  
\en
\be
z_{12}=(\bar E_1 \Delta x_1 + \bar E_2 \Delta x_2)/E_{\bar p},  
\en
where  $\Delta x_i=x_i-\bar v_i t_i$ for $i=1,2$. 
For $t_1=t_2=t$ we have $t_{12}=t-(x_1-x_2)/\bar v_{12}$  and $z_{12}=x-\bar v t$,
where $x=(\bar E_1 x_1 + \bar E_2 x_2)/E_{\bar p}$ is the center of mass of the two
final state particles. Hence $z_{12}$ is the deviation of the center of mass from classical motion, and
$t_{12}=t_d$ corresponds to the stochastic time $t_d$  at which the particle pair is created
at  $x_1=x_2 = x_d$,  where  $\phi_{12}=\bar p x_d -E_{\bar p} t_d$ and
$z_{12}=x_d-\bar v t_d$. At  such a point the final state  wavefunction
is proportional to the initial wavefunction at $x=x_d,t=t_d$, and it is 
independent of the mass of the decay  particles.  Consequently this wavefunction satisfies  
the initial  condition that the flavor of the state be independent 
of $x_d,t_d$,  an important result which could  
not be imposed  ab initio.  As expected, for $ t-t_{12} < 0 $, this  wavefunction vanishes.

In our approximation, the initial function, Eq. \ref {init1},  
determines directly the envelope $g(z_{12})$ of the wave packet of the 
two  final state  particles.  
Hence the  probability for finding these particles at  $x_1,t_1$ and $ x_2,t_2 $  is given by
\be
\label{envelope2}
\frac {\Gamma M}{E_{\bar p} \bar v_{12}} g^2(z_{12})~ exp(-\Gamma M t_{12}/E_{\bar p}) dx_1 dx_2  
\en
It can be readily  verified that the probability for the creation of these 
two particles at $x_d,t_d$ is equal to the probability that the source decays 
at this  same space-time point. If this decay is not measured or constrained by 
the environment,  then Eq. \ref {envelope2} can be applied directly to calculate 
probabilities or averages over the recoil variable coordinates.
However if $t_d$ is observed 
then for times  $t_1 \geq t_d$ and $t_2 \geq t_d$  the probability for finding the particles
at $x_1,x_2$ is  obtained by  setting  $t_{12}=t_d$ in Eq. \ref{envelope2}  (footnote 2), and 
$ x_1-x_2= \bar v_1 (t_1-t_d)-\bar v_2 (t_2-t_d)$, or
\be
\label {deld}
\Delta_d x_1=\Delta_d x_2=z_{12}
\en
where $\Delta_d x_i=x_i-\bar v t_d -\bar v_i (t_i-t_d)$ is the deviation from classical motion
of the decay  particles. In this case, the measure $z_{12}$ for this deviation satisfies
he same distribution as the deviation from classical motion of the initial state,
and therefore our  analysis shows that
particles associated with the decay process are confined to 
move along  classical trajectories  with the same degree of localization
as the source.

We now assume that the two states labeled $a$ and $b$  correspond 
to particles $1$ and $2$ with small mass differences, $\del m_i^2 = m_{ia}^2-m_{ib}^2$,   
and calculate the corresponding
differences in the   mean momentum $\bar p_1$ and $\bar p_2$
and corresponding energies $\bar E_1$ and $\bar E_2$  from the energy-momentum
conservation laws, Eqs. \ref {mom1} and  \ref{en1}. We have  
\be
\label{delp1}
\del \bar p_1 +\del \bar p_2 =0,
\en
and 
\be
\label {delE1}
\del \bar E_1 +\del \bar E_2 =0,
\en
where to first order in 
$\del m_i^2 $,
\be
\label{ener1}
\del \bar E_i= \bar v_i \del \bar  p_i + \frac {\del m_i^2}{2 \bar E_i}.
\en
Solving these equations we obtain 
\be
\label{delp2}
\del \bar p_1 = -\del \bar p_2 =  -\frac {1}{ \bar v_{12}} (\frac {\del m_1^2}{2 \bar E_1} 
+\frac{\del m_2^2}{2 \bar E_2}).
\en
where $\bar v_{12}=\bar v_1 -\bar v_2$ is the relative velocity. 
These relations  differ from the result obtained with  conventional 
kinematics  assumptions  that different mass states  have either the same  
momentum \cite {nussinov1} - \cite {kayser1}  or the same energy \cite {lipkin1} -\cite {lipkin4}. 

The oscillation term which concerns us here appears  in the calculation of the 
interference term in the transition probability $ |A|^2$, where $A$ is
given by  Eq. \ref {amplitude1}, and  is proportional to 
\be
\label{interf1}
Real \psi_a^{*} \psi_b= cos (\phi) g(z_{12,a}) g(z_{12,b})exp(-M\Gamma (t_{12,a}+t_{12,b})/2E_{\bar p})
\en
where we have ignored factors which  depend  on the lifetime of the final state particles,
Eq. \ref {lifetimes1}.  The phase difference 
$\phi=\phi_{12,a}-\phi_{12,b}=\del \bar p_1 x_1 +\del \bar p_2  x_2 -\del \bar E_1 t_1 -\del \bar E_2 t_2$
is invariant under Lorentz transformations, and according to the energy-momentum
conservation laws, Eqs. \ref {delp1} and \ref {delE1}, it can be written in the form
\be
\phi= \del \bar p_1 (x_1-x_2)-\del \bar E_1 (t_1-t_2).
\en
This form shows that the phase difference $\phi$  depends only on the relative time coordinate of the final
state particles and, therefore, that it is  independent of the initial decay point  $t_d$.
However the role of this decay time  appears when we consider the effect of
the wave packet envelope in the case that this coordinate is measured,
constraining each of the decay  particles to move 
near its classical trajectories. For this purpose we apply
Eq. \ref {ener1}
to write $\phi$ in the equivalent form
\be
\label{phifinal3}
\phi=  \del \bar p_1(\Delta x_1-\Delta x_2) -\frac{\del m_1^2}{2 \bar E_1} t_1
-\frac{\del m_2^2}{2 \bar E_2}t_2,
\en
This expression for  $\phi$  is  similar to  results 
given in \cite {okun1} and  \cite {goldman2}, but our correlated
wave packet now allows us  to interpret  and evaluate properly  the contribution of the 
first term in $\phi$ which does not appear in the convential single
particle formulation for this  phase. 
Substituting Eq. \ref {delp2} for $\delta \bar p_1$
and substituting  $\Delta x_2-\Delta x_1 =  
\bar v_{12} t_{12}$, we obtain finally
\be
\label{phifinal1}
\phi=  -\frac{\del m_1^2}{2 \bar E_1} (t_1-t_{12}) 
-\frac{\del m_2^2}{2 \bar E_2}(t_2-t_{12}),
\en
If the recoil particle has a fixed mass, i.e. $\del m_2^2=0$, this 
form for $\phi$  is equal  to the conventional single particle  result 
with mass eigenstates of the {\it same momentum} provided that we identify $t_{12}$
with the decay  time  $t_d$  of the initial state as we have done previously,
see footnote 2. In this case this phase is independent of the
recoil particle coordinates contrary to assertions in \cite {widom0}-\cite {widom3}. 
However, if the decay time $t_d$ is not measured  directly,  then
in principle it could be determined from a coincidence measurement
of  the recoil coordinate $x_2$.
If we set $t_1=t_2$, and substitute  $t_1-t_{12}=(x_1-x_2)/\bar v_{12}$ in Eq. \ref {phifinal1}, we obtain 
EPR-like oscillations in the relative coordinates of the two final state particles 
which can only  be observed if the flavor is also determined \cite {okun1}.
In the case of  neutral meson pairs in the final state, e. g.  $ K \bar  K$  
produced in $\phi$  decay, $\del m_2^2=-\del m_1^2$, and
the time  coordinates of both particles appear in Eq. \ref {phifinal1}
even if the decay time $t_d$ has been determined. 
A similar  result was  obtained in  \cite {kayser2} and \cite {rosner}  by assuming 
that $\del \bar p_1 = \del \bar p_2 =0 $, although 
this  kinematical condition is not justified.

If neither the decay time $t_d$ nor the 
coordinates $x_2,t_2$ of the recoil particle  are observed,
we must integrate  the interference term of the transition probability, 
Eq. \ref {interf1}  over the decay time $t_d$, or what amounts to an  
{\it equivalent procedure},  over the position coordinate $x_2$  of
the  unobserved recoil particle.
Assuming that there are no enviromental constraints on the possible range
of these variable, we apply eq. \ref {envelope2} 
to obtain the average of $cos(\phi)$.
Neglecting the mass difference in the envelopes of the wave packet,
which leads to a finite coherence  length, we obtain 
\be
\label{avcost}
<cos(\phi)>=R ~cos~(\frac{\del m_1^2}{2 \bar E_1}( t_1-t_s) + \frac {\del m_2^2}{2 \bar E_2}(t_2-t_s) -\del),
\en
where
\be
R=\frac{1}{\sqrt{1+\xi^2}},
\en
\be
tan(~\del)= \xi,
\en
and
\be
\xi= (\frac {\del m_1^2}{2 \bar E_1}+\frac {\del m_2^2}{2 \bar E_2})(\frac{\bar E_p}{M\Gamma}).
\en
This average is independent of the shape of the initial wave packet.
The parameter $\xi$ gives a measure for the deviation from the conventional form, which 
corresponds to  $R=1$ and $\del=0$ in Eq. \ref {avcost}. 
For example, in  particle reactions producing neutral mesons, $\Gamma$ is of order several {\it MeV}, 
and for the B meson  $\delta m$ is $ 3.1 \times 10^{-4}  {\it eV} $ , and about 
100 times smaller  for the K meson. Hence  
$\xi$ is of order $ 10^{-10} -10^{-12}$, and
the  contribution from the first term which  appears in Eq. \ref {phifinal3}
is  essentially  unobservable, contrary to
expectations  in \cite {goldman2}.
For neutrinos produced in pion decay,  with
$\del m^2 \approx 10^{-3} eV^2$  
as found in \cite {kamiokande1},   $\xi \approx  10^{-3} $ 
can also be neglected in the analysis of the data. However, 
this is not the case for  $\del m^2 $ of order
$1-10$ $ eV^2 $ as reported in  \cite {lsnd2}.  Furthermore, the magnitude of $\xi$ is 
greater for the case that the neutrinos are produced in muon decay \cite {lsnd1}
because the muon  is lighter than the pion and has a longer life time, 
but this is a three body decay and our analysis is only approximatly valid.

In practice it must be remembered that oscillations
are observed in position rather than in  time  coordinates. Setting
$t_i=(x_i-\Delta x_i)/\bar v_i$ in Eq. \ref {phifinal1} for the phase  we obtain
an equivalent form
\be
\label{phifinal2}
\phi= (\frac {\del m_1^2}{2 p_1}+\frac {\del m_2^2}{2p_2}) z_{12}
-\frac{\del m_1^2}{2 \bar p_1} (x_1-\bar v t_{12}) 
-\frac{\del m_2^2}{2 \bar p_2}(x_2-\bar v t_{12}).
\en
This differs from the conventional form for $\phi$, but 
if the decay coordinates are determined 
then  $t_{12}=t_d$, $z_{12}=\Delta_d x_1=\Delta_d x_2 $, Eq. \ref {deld},  and we recover the conventional form.
Notice that in the rest frame of the source $\bar v =0$, and the dependence of
the phase on $t_{12}$  vanishes because  the source is not moving.
Assuming a Gaussian distribution for the initial wave packet with a width $\sigma_x$, and averaging
$cos(\phi)$ over $z_{12}$ and $t_{12}$ we now obtain
\be
\label{avcosx}
<cos(\phi)>=\bar R ~cos~(\frac{\del m_1^2}{2 \bar p_1}(x_1-x_s) + \frac {\del m_2^2}{2 \bar p_2}(x_2 -x_s) - \bar \del),
\en
where
\be
\bar R=\frac{exp(-\eta^2) }{\sqrt{1+\bar \xi^2}},
\en
\be
tan(~\bar \del)= \bar \xi,
\en
\be
\bar \xi  = (\frac {\del m_1^2}{2 \bar p_1}+\frac {\del m_2^2}{2 \bar p_2})(\frac{\bar p}{M\Gamma}),
\en
and 
\be
\bar \eta = (\frac {\del m_1^2}{2 \bar p_1}+\frac {\del m_2^2}{2 \bar p_2}) \sigma_x.
\en
As expected from simple physical arguments, interference effects can occur in position measurements provided that the 
width $\sigma_x$ of the wave packet is small
compared to the oscillation length $ \propto  \bar p_i/\del m_i^2$,
or correspondingly that  $\eta \propto \del \bar p_i \sigma_x << 1$. 
The magnitude of  $\sigma_x$ is of the order of magnitude of the localization  
of a nuclear target, and it is further contracted by the Lorentz transformation
due to the motion of the unstable initial particle  so that in practice
$\eta <<1$.
Moreover, in the rest frame of the source $\bar p = 0 $ and consequently
in this case $\bar \xi = 0$.

In conclusion, we have shown that the  transition probability  
for neutrino and neutral meson oscillations
can be  obtained from first principles by solving the time dependent 
Schr\"{o}dinger equation for the decay of an unstable source
into a coherent superposition of  correlated two particle eigenstates
with different masses. We have obtain our results in a relativistically
covariant manner by applying  well defined approximations without recourse to  
conventional ad-hoc assumptions which violate principles of quantum mechanics, 
and have led to much confusion in the literature. 
We have shown that the  width or lifetime of the source plays a crucial role in 
understanding this problem, and  that
quantum  correlations between the final state particles 
relate the decay time of the source
to the space-time coordinates of the these particles.
Due to the mass difference the wave packets for different mass eigenstates 
have different group velocities and separate leading to a 
finite coherence length \cite {giunti1}, 
but this effect  was  neglected  here. 

\subsection*{Acknowledgements}

I would like to thank D. Dorfan and  A. Seiden for very stimulating  discussions,
critical questions, and a careful reading of this manuscript, 
and T. Goldman and Ann Nelson for several useful comments.

\subsection*{Footnotes}

\begin{enumerate}

\item The justification for Eq. \ref {psi11}  is that after one of the two particles
has been  detected, i.e. $t=t_1$ or $t=t_2$, its state  does not continue to evolve in time. 
The  generalized  wavefunction $\psi (x_1,x_2,t_1,t_2)$, 
Eq. \ref {psi12} can then be  interpreted as  
the probability amplitude for correlated  events which occur at the two different
space-time points   $(x_1,t_1)$ and $(x_2,t_2)$, and  is equivalent  to the  amplitude method in  
\cite {kayser2}, and the formalism in \cite {okun1}.  
It can be shown that this procedure is  equivalent  to the  ``collapse'' of the wavefunction
language, which is the conventional description  when measurements take place at different times,
although it is preferable not to invoke this awkward language. 

\item  For $t_1,t_2 \geq t_d$, the condition  $t_{12} = (\bar v_1 t_1 - \bar v_2 t_2 -x_1 +x_2)/ \bar v_{12} = t_d$ 
connects the decay time $t_d$ of the source
to the space-time coordinates of the decay  particles. 
It corresponds to the  {\it  classical} relation for the relative
coordinates of these particles which can be understood 
on the physical grounds that these  particles are created in a region of negligible small 
spatial dimension, without  violating  the uncertainty principle
because $t_d$ is a stochastic variable. This relation has also  been obtained in \cite {okun1},
but under the invalid assumption that the source and decay particles follow  classical trajectories 
exactly. In the case that the decay can  occur
over a range of values $ 0\leq t_d \leq t_{max}$, one must take an average 
over the probability distribution integrated over
this range provide  there are no measurements on the recoil particle  
which constrain the possible  values of $t_{12}$. 
In practice  $t_{max}= d / \bar v$, where $d$  is the distance 
between the  target
where the unstable particle is created, and a beam stop where nuclear reactions 
annihilate it.

\end{enumerate}

\end{document}